# Plasmonic and photonic enhancement of chiral near fields

*Li Hu[1, #], Zhiguang Sun[2, #], Yingdong Nie[2], Yingzhou Huang[3], and Yurui Fang[2, *]*

[#]These authors contributed equally.

L. Hu

Chongqing Engineering Laboratory for Detection, Control and Integrated System, School of Computer Science and Information Engineering, Chongqing Technology and Business University, Chongqing, 400067, P. R. China

Z. Sun, Y. Nie, Y. Fang

Key Laboratory of Materials Modification by Laser, Electron, and Ion Beams (Ministry of Education); School of Physics, Dalian University of Technology, Dalian 116024, P.R. China
E-mail: yrfang@dlut.edu.cn

Y. Huang

State Key Laboratory of Coal Mine Disaster Dynamics and Control, Chongqing Key Laboratory of Soft Condensed Matter Physics and Smart Materials, College of Physics, Chongqing University, Chongqing 400044, China

A chiral near field with a highly contorted electromagnetic field builds a bridge to match the chiral molecules and light wavelengths with large size differences. It significantly enhances the circular dichroism of chiral molecules and has great prospects in chirality sensing, detection, trapping, and other chirality-related applications. Surface plasmons feature outstanding light-trapping and electromagnetic-field-concentrating abilities. Plasmonic chiral nanostructures facilitate light manipulation to generate superchiral near fields. Meanwhile, the nanophotonic structures have attracted significant interest to obtain strong chiral fields due to their unique electromagnetic resonant properties. During the interaction of light and chiral materials, the chiral near field not only bridges the light and chiral molecules but is also responsible for the optical activities. This paper reviews state-of-the-art studies on chiral near field enhancement using plasmonic and photonic nanostructures. We review the principle of chiral electromagnetic fields and the development of plasmonic and photonic nanostructures for near field enhancement. The properties and applications of enhanced chiral near fields for chiral molecule detection, spin-orbit angular interaction, and the generation of the chiral optical force are examined. Finally, we discuss current challenges and provide a brief outlook of this field.

**1. Introduction**

Chirality refers to the symmetrical nature of objects. A chiral object has a three-dimensional (3D) structure that cannot be superimposed on its mirror image by rotation or translation, e.g., human hands. Objects with opposite chirality are called enantiomers. They are identical in chemical composition and have the same scalar physical properties, i.e., density, enthalpy of formation, and vibrational frequency. Their differences are only observable when they interact with other chiral objects.[1] Chiral objects are ubiquitous in nature, ranging from sub-atomic particles to galaxies. Chirality is also essential to human lives because most biomolecules, the building blocks of life, including amino acids, nucleic acids, and carbohydrates, exist only in one chiral configuration.[2,3] Thus, enantioselective reactions occur in living organisms.[4] Thalidomide is a good example of the different effects of enantiomers in the body. It is harmless as a painkiller for pregnant women, but its enantiomer causes malformation of infants.[5]

The absorption difference, i.e., circular dichroism (CD), is widely used to distinguish the chiral states of molecules based on the different interactions of molecular enantiomers with left-hand and right-hand circularly polarized light (CPL), which is a chiral electromagnetic field.[6] However, the CD of small chiral molecules is very weak because the absorption cross-sections of left and right CPL differ by less than one in a thousand. This is a result of the size mismatch of the molecules and the light wavelength. At the scale of small molecules, the CPL field exhibits only a small change in the helical pitch with a slight twist, resulting in weak excitation perturbation.[7]

Two strategies have been used to enhance the CD intensity. The first focuses on the molecular aspect of CD and designs molecules with large optical dissymmetry. The second uses exciting electromagnetic fields that are more contorted than commonly used CPL fields.[7] Tang and Cohen proposed to enhance the field dissymmetry by shortening the field line reorientation distance relative to the free-space wavelength. They were the first authors to propose the chiral near field in 2010 to obtain "superchiral light" with a highly distorted electromagnetic field.[1] They obtained a light field with chiral asymmetry much greater than that found in CPL plane waves at the nodes of an optical standing wave. The authors conducted an experiment and observed an 11-fold enhancement over CPL in the discrimination of the enantiomers of a biperylene derivative.[7]

Localized surface plasmons (LSPs) exhibit outstanding light-trapping and electromagnetic-field-concentrating abilities.[8] Therefore, chiral plasmonic nanostructures were proposed as an effective and reliable approach to obtain a chiral near field in the vicinity of the nanostructures.[9] The principles for designing chiral plasmonic nanostructures with strong chiral near fields were proposed.[10] In recent years, different types of plasmonic nanostructures have been designed to generate enhanced chiral electromagnetic near fields, ranging from planar to 3D, monomer to polymer, homogeneous to heterogeneous, and chiral to achiral structures. The excitation light was also expanded from CPL to linearly polarized light.[11-16] Similarly, some nanophotonic structures have been proposed to enhance chiral fields due to their strong electromagnetic responses and low loss.[17-21] The enhanced chiral near field based on well-developed nanofabrication techniques exhibits superior convenience and dissymmetry enhancement. This approach shows great prospects for chirality sensing, detection, trapping, and other chirality-related applications.

Here, we review publications on the plasmonic and photonic enhancement of chiral near-fields. First, we introduce the optical principle of chirality and the related theory. Second, the design and fabrications of nanostructures with enhanced chiral near-fields are discussed, including 3D, planar two-dimensional, and achiral nanostructures, handed and uniform chiral near-fields, and metal and dielectric structures. We describe the applications of chiral near-fields in chirality sensing, detection, and trapping. Finally, we summarize the reviewed studies and provide a future outlook of this field.

## 2. Principle and theory

Helicity is a commonly used parameter to describe chiral fields. The helicity of any divergenceless vector field $\boldsymbol{B}(\boldsymbol{r})$ in a domain $\boldsymbol{D} \subset R^3$ can be obtained by the integral[22]

$$h(\boldsymbol{B}) = \int_{\boldsymbol{D}} \boldsymbol{A} \cdot \boldsymbol{B} d^3 r \tag{1}$$

where $\boldsymbol{A}$ is a vector potential of $\boldsymbol{B}$. Since $\boldsymbol{B}$ is the curl of the vector potential $\boldsymbol{A}$ and defines the rotation of $\boldsymbol{A}$ around a point, helicity indicates how much $\boldsymbol{A}$ rotates around itself times its modulus. In an electromagnetic field in a vacuum, the electric field $\boldsymbol{E}$ and magnetic field $\boldsymbol{B}$ are divergenceless, and their vector potentials $\boldsymbol{A}$ and $\boldsymbol{C}$ are:

$$\boldsymbol{E} = \nabla \times \boldsymbol{C} = -\frac{\partial \boldsymbol{A}}{\partial t} \tag{2}$$

$$\boldsymbol{B} = \nabla \times \boldsymbol{A} = -\frac{\partial \boldsymbol{C}}{\partial t} \tag{3}$$

The helicity of the electromagnetic field is defined as:[23]

$$\mathcal{H} = \frac{1}{2}\int (\boldsymbol{A}\cdot\boldsymbol{B} - \boldsymbol{C}\cdot\boldsymbol{E})\, d^3r \tag{4}$$

Here, the first and second terms are the magnetic field helicity ($\mathcal{H}_{mag}$) and the electric field helicity ($\mathcal{H}_{elc}$), respectively[22]

$$\mathcal{H}_{elc} = \frac{1}{2}\int (\boldsymbol{C}\cdot\boldsymbol{E})\, d^3r \tag{5}$$

$$\mathcal{H}_{mag} = \frac{1}{2}\int (\boldsymbol{A}\cdot\boldsymbol{B})\, d^3r \tag{6}$$

$\mathcal{H}$ is a conserved *pseudoscalar*, and it is gauge-invariant under the same conditions as $\mathcal{H}_{mag}$ and $\mathcal{H}_{elc}$.

From another aspect of particle physics, helicity is also defined as the projection of the angular momentum in the motion direction:

$$\mathcal{H}' = \frac{\boldsymbol{p}\cdot\boldsymbol{J}}{p} \tag{7}$$

where $\boldsymbol{p}$ is the momentum, and $\boldsymbol{J}$ is the angular momentum, which is defined as[22, 23]

$$\boldsymbol{J} = \varepsilon_0 \int \boldsymbol{r}\times(\boldsymbol{E}\times\boldsymbol{B})\, d^3r \tag{8}$$

It is known that $\boldsymbol{J}$ can be separated into two parts, i.e., the orbital part ($\boldsymbol{L}$) and the spin part ($\boldsymbol{S}$).[24, 25]

$$\boldsymbol{L} = \varepsilon_0 \hat{r}_i \int d^3r[E_\alpha(\boldsymbol{r}\times\boldsymbol{\nabla})_i A_\alpha] \tag{9}$$

$$\boldsymbol{S} = \varepsilon_0 \hat{r}_i \int d^3r(\boldsymbol{E}\times\boldsymbol{A})_i \tag{10}$$

The spin angular momentum of chiral light fields is associated with circular polarization (CP), and the orbital angular momentum depends on the presence of helical phase fronts.[23]

In 1964, Lipkin introduced a conserved quantity of 00-zilch:[26]

$$\mathcal{Z}^{00} = \frac{1}{2}\int (\boldsymbol{E}\cdot\boldsymbol{\nabla}\times\boldsymbol{E} + \boldsymbol{B}\cdot\boldsymbol{\nabla}\times\boldsymbol{B})d^3r \tag{11}$$

It is a higher-order extension of the optical helicity obtainable from the optical helicity by replacing the electric and magnetic fields with their curls.[23] However, Lipkin and others subsequently dismissed this quantity as having no physical significance until Tang et al. employed it to measure the chirality of light. This quantity is known as the optical chirality or the chirality density, i.e., geometrically, the field lines wrap around a central axis with a component parallel to that axis.[1] Similar to the energy density and Poynting energy flow, the chirality density ($C$) and the chirality flow ($\phi$) are expressed as:

$$C = \frac{\varepsilon_0}{2}\boldsymbol{E}\cdot\boldsymbol{\nabla}\times\boldsymbol{E} + \frac{1}{2\mu_0}\boldsymbol{B}\cdot\boldsymbol{\nabla}\times\boldsymbol{B} = -\frac{\varepsilon_0\omega}{2}\mathrm{Im}(\mathbf{E}^*\cdot\mathbf{B}) \tag{12}$$

$$\phi = \frac{\varepsilon_0 c^2}{2}[\boldsymbol{E}\times(\boldsymbol{\nabla}\times\boldsymbol{B}) - \boldsymbol{B}\times(\boldsymbol{\nabla}\times\boldsymbol{E})] \tag{13}$$

which also satisfy the continuity equation:[12, 27]

$$\frac{\partial C}{\partial t} + \nabla \cdot \phi = 0 \tag{14}$$

The interaction between chiral molecules and an electromagnetic field is described by the excitation rate. When a chiral molecule is subjected to a monochromatic electromagnetic field, it generates an electric dipole moment **p** and a magnetic dipole moment **m** defined as:

$$\widetilde{\mathbf{p}} = \tilde{\alpha}\widetilde{\boldsymbol{E}} - i\tilde{G}\widetilde{\boldsymbol{B}}, \ \widetilde{\mathbf{m}} = \tilde{\chi}\widetilde{\boldsymbol{B}} + i\tilde{G}\widetilde{\boldsymbol{E}} \tag{15}$$

Where $\tilde{\alpha} = \alpha' + i\alpha''$ is the electric polarizability, $\tilde{\chi} = \chi' + i\chi''$ is the magnetic susceptibility, and $\tilde{G} = G' + iG''$ is the isotropic mixed electric-magnetic dipole polarizability. The excitation rate of the molecule by the field is[1]

$$A^{\pm} = \langle \boldsymbol{E} \cdot \dot{\mathbf{p}} + \boldsymbol{B} \cdot \dot{\mathbf{m}} \rangle = \frac{\omega}{2}(\alpha''|\widetilde{\boldsymbol{E}}|^2 + \chi''|\widetilde{\boldsymbol{B}}|^2) \pm G''\omega Im(\widetilde{\boldsymbol{E}}^* \cdot \widetilde{\boldsymbol{B}}) \tag{16}$$

where the brackets indicate an average over time. Since $\chi$ is small for most molecules, the second term of equation (16) can be ignored. By substituting $\omega Im(\widetilde{\boldsymbol{E}}^* \cdot \widetilde{\boldsymbol{B}}) = \dot{\boldsymbol{B}} \cdot \boldsymbol{E} - \dot{\boldsymbol{E}} \cdot \boldsymbol{B}$, the following is obtained:

$$A^{\pm} = \frac{2}{\epsilon_0}(\omega U_e \alpha'' \mp C G'') \tag{17}$$

where $U_e = \frac{\epsilon_0}{4}|\widetilde{E}|^2$ is the time-averaged electric energy density. The dissymmetry factor is calculated by $g = 2(A^+ - A^-)/(A^+ + A^-)$; thus, $g$ includes any pair of electromagnetic fields defined as[1]

$$g = -\left(\frac{G''}{\alpha''}\right)\left(\frac{2C}{\omega U_e}\right) \tag{18}$$

Here, the first term in equation (18) denotes the parameters of the chiral molecules, and the second term denotes those of the electromagnetic fields. The equation indicates that the chiral asymmetry of the excitation rate of the chiral molecules is proportional to the product of the chirality of the molecules and the electromagnetic fields. Large chiral asymmetry can be achieved by constructing a superchiral near field with a large value of the second term in equation (18). It is known that the dissymmetry factor excited by CPL is $g_{CPL} = -4G''/c\alpha''$; thus, the increase in the dissymmetry factor by using a superchiral near field compared to CPL can be evaluated by:

$$\hat{g} = \frac{g}{g_{CPL}} = \frac{cC}{2\omega U_e} \tag{19}$$

## 3. Plasmonic chiral fields in different structures

Since the concept of the "superchiral field" and its quantitative description were proposed,[1] researchers have developed various plasmonic nanostructures to obtain strong near-field chiral responses due to local electromagnetic fields.[7, 9] Research has shown that the nanostructure configuration and the incident light characteristics substantially affect the enhancement of chiral fields.

**3.1. Chiral near fields in the vicinity of 3D structures**

As the most intuitive chiral structure, helical structures have attracted much attention for generating chiral near fields. When the rotational direction of a single helix matches the polarization of incident CPL, chiral near-field responses occur, as shown in **Fig. 1**a. Meanwhile, the enhanced optical chirality of its enantiomers is symmetric.[10] For multiple helices, the chiral electromagnetic fields over an extended region are illuminated by the linearly polarized light, which is just used to stimulate the resonance of the structure. The single-handed chiral near-fields in the helices are the results of the chiral eigenmodes of the structures, which could be used for discriminating enantiomers.[11] Besides helical nanostructures, 3D chiral oligomers composed of nanoparticles have drawn much interest due to the tunability and strong coupling effect between particles in the generation of superchiral fields.[10, 28, 29] In 2012, Schäferling et al. proposed a chiral oligomer composed of two twisted L-shaped layers.[10] Similar to the helix, strong chiral near-field responses occur with matching polarized light illumination. The strongest chirality enhancement is observed in the small gaps between two disks due to near-field enhancement. In 2014, Ogier et al. fabricated chiral nanoparticle oligomers (Fig. 1b) with facile hole-mask colloidal lithography to achieve strong chiral far-field and near-field responses.[28] Chiral hotspots with strong chiral near fields ($\frac{C}{|C_{CP}|} \sim 200 - 300$) occur in the gap between nanoparticles, corresponding to the "hot spots" of electromagnetic fields. The average optical chirality surrounding the oligomers is ~5 at the plasmonic resonance peak. This method may have potential for the detection and analysis of small number of molecules.

**3.2. Chiral near fields in the vicinity of 2D structures**

However, the preparation of 3D chiral structures is relatively complex. It has been demonstrated that planar chiral structures could be used to generate chiral fields. [9, 10, 30-33] The gammadion structure, a classic planar chiral structure, has been widely studied theoretically and experimentally.[9, 10, 32] In 2012, Schäferling et al. demonstrated that the regions of the highest enhancement of the chiral fields and electric fields in the gammadion structure are not identical, indicating that strong electric fields are not sufficient to produce superchiral fields.[10] Unlike the helices, the superchiral field of the grammadion is insensitive to the polarization state of incident CPL and is discontinuous in spaces. To overcome these disadvantages, a two-armed nanospiral was proposed to improve the chiral near-field properties (Fig. 1c).[10] As expected, the chiral near-fields of the nanospirals are different for different incident polarizations, and two opposite chiral fields occur on either side of the structure. This approach could have potential for enantiomer sensing and optical rotation measurements. In 2017, Kang et al. achieved the precise control of chiral near-fields using "superchiral light",[7] including chiral switching and selective enhancement.[32] In addition, the superchiral fields are more delocalized, which improves the potential for nanostructure-based enantiomeric sensing. Furthermore, simpler structures such as staggered nanorod dimer[30] and nanoslits[31] arrays were theoretically and experimentally studied. Hendry et al. investigated the physical mechanism of chiral near-fields and demonstrated the generation of chiral evanescent fields with nanoslits arrays based on the modal matching theory.[31] As shown in Fig. 1d, when the electric field of one slit overlapped with the magnetic field of the other slit with a π/2 phase difference, strong chiral near-field responses are observed. The handedness of chiral near-fields depends on the structure rather than on the polarization of incident light. However, the chiral near-field rapidly decays with the distance from the surface, and the handedness of the two sides of the array is just opposite.

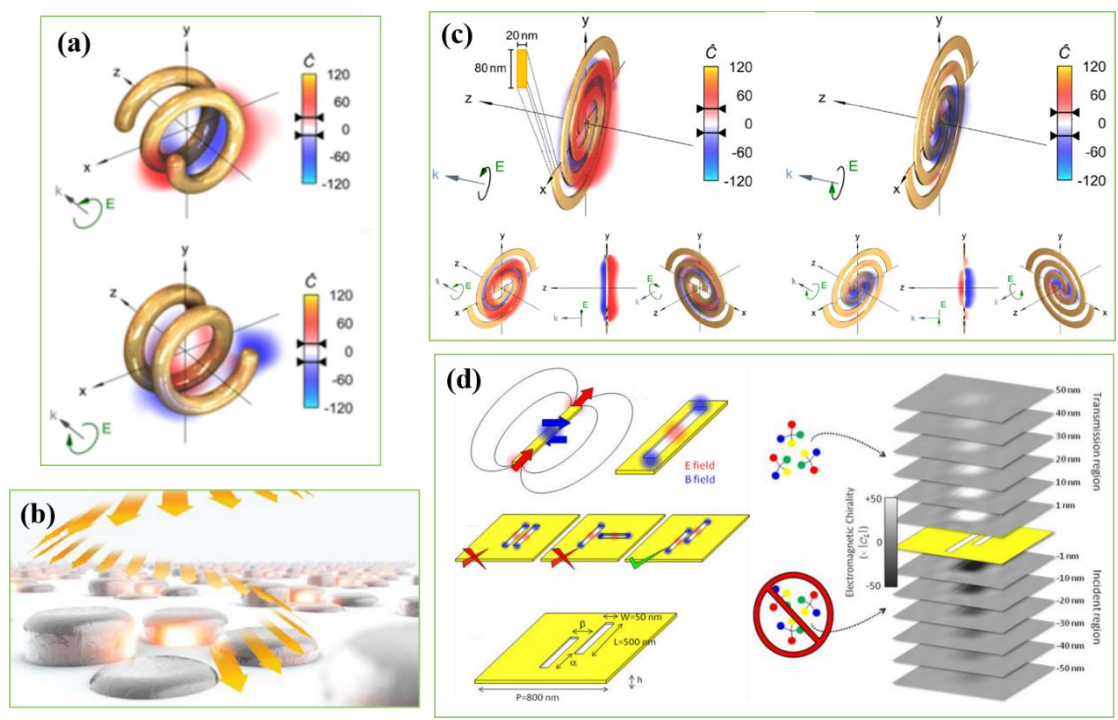

**Figure 1.** (a) Optical chirality enhancement of helixes excited by matching handed circularly polarized light (CPL).[10] This figure has been adapted from ref. 10 with permission from American Physical Society, copyright 2012, under the Creative Commons Attribution 3.0 License. (b) Schematic of chiral nanoparticle oligomers excited by CPL.[28] This figure has been adapted from ref. 28 with permission from American Chemical Society, copyright 2014. (c) Optical chirality enhancement of planar chiral nanostructure under left-handed CPL (LCP) and right-handed CPL (RCP) illuminations.[10] This figure has been adapted from ref. 10 with permission from American Physical Society, copyright 2012, under the Creative Commons Attribution 3.0 License. (d) Schematic of chiral near fields generated with a nanoslit dimer and the chiral near-field distributions.[31] This figure has been adapted from ref. 31 with permission from American Chemical Society, copyright 2012.

**3.3. Chiral near fields in the vicinity of nonchiral structures with linear polarization**

Similar to the far-field chiral responses of plasmonic nanostructures, local chiral fields have been observed near symmetric nanostructures under certain conditions.[12, 13, 34, 35] This response is called "extrinsic chirality". Different from chiral near fields of chiral nanostrcutuers excited by CPL, the Schäferling group and Davis group theoretically demonstrated that superchiral fields could be generated with achiral nanostructures under linearly polarized light.[12, 34] As shown in **Fig. 2**a, chiral near-fields appear near a nanoantenna under normal incidence and are caused by the interference between incident and scattered fields.[12] The authors explained the phenomenon using a simple dipole model. Similarly, Davis and Hendry obtained chiral near-field enhancement using an achiral structure composed of three nanorods.[34] As discussed above, uniform and strong superchiral fields are very important for multiple applications. Tian et al. designed a simple Au dimer structure and excited it with linearly polarized light in the diagonal direction of the structure,[13] as depicted in Fig. 2b. Simulation results showed that one-handed chiral near-fields and electric fields are significantly enhanced in the gap of dimers, and the handedness of the chiral fields could be inverted by changing the incident polarization to the other diagonal direction of the dimer. The volume-averaged chiral field enhancement in the whole gaps is 30-fold, exhibiting the potential for chiral sensing and Raman optical activity (ROA) measurements of small quantities of chiral molecules. The authors explained the mechanisms using an analytical dipole model. On this basis, Hu et al. enhanced the local chiral field by adding an Au film under the block dimer.[36] Further, they used a block heterodimer with different sizes on the Au film and achieved one-handed chiral fields with opposite chirality in different regions.[37]

**3.4. Hidden chirality**

Different from "extrinsic chirality" with off-resonance, "hidden chirality" was obtained in achiral nanosystems with symmetrical stimulations[38-40] resulting from the interference between the two plasmonic eigenmodes excited by CPL. In 2018, Zu et al. acquired an image and controlled the chiral electromagnetic modes of the V-shaped nanostructure with CP-with cathodoluminescence CL microscopy (Fig. 2c).[38] The chiral electromagnetic modes and chiral radiative local density of nanostructures' states were observed in the CL image. The chirality of the symmetric nanostructure has an extreme chiral distribution (~99%) at the arm ends, which can be used to detect polarized CL emission from the hotspot and determine the radiative local density of states (LDOS). Subsequently, Horrer et al. achieved symmetric local optical chirality with a symmetric trimer consisting of three gold nanodisks[39] under LCP and RCP, as shown in Fig. 2d. They experimentally demonstrated that the optical chirality could be imprinted onto the photosensitive polymer, which could be used in polarization-sensitive photochemistry.

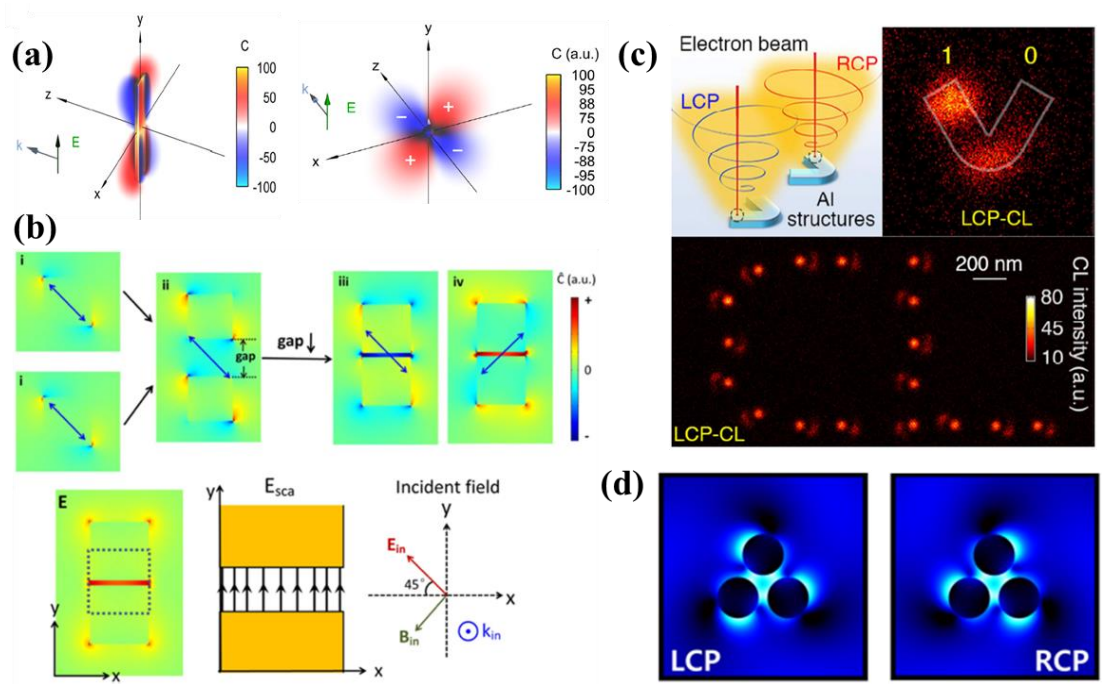

**Figure 2.** (a) Chiral near-field distributions of a nanoantenna and a dipole excited by linearly polarized light.[12] This figure has been adapted from ref. 12 with permission from The Optical Society, copyright 2012. (b) Schematic of the formation of uniform chiral near-fields in the gap of a gold block dimer.[13] This figure has been adapted from ref. 13 with permission from Nature Publishing Group, copyright 2015. (c) Schematic of V-shaped units illuminated by LCP and RCP and the LCP CL image.[38] This figure has been adapted from ref. 38 with permission from American Chemical Society, copyright 2018. (d) Chiral near-fields distributions of plasmonic trimer excited by LCP and RCP.[39] This figure has been adapted from ref. 39 with permission from American Chemical Society, copyright 2020.

## 4. Chiral near fields generated by photonic enhancement

As discussed above, various plasmonic nanostructures composed of metallic elements have been designed to obtain strong chiral fields due to highly enhanced local electromagnetic fields. Equation (12) indicates the importance of the magnetic field to generate chiral fields. However, the magnetic modes of metal structures are either too weak or separate from the electric modes, resulting in challenges to achieve strong and one-handed chiral fields. Some negative-index metamaterials and all-dielectric nanostructures have been utilized to improve the design and obtain chiral fields.

Since negative-index materials can generate strong local magnetic fields, Yoo et al. proposed an achiral negative-index metamaterial composed of stacked double-fishnet layers (**Fig. 3**a) to obtain highly enhanced chiral fields.[41]

Large-volume and one-handed chiral fields are generated in entire volume inside each cavity, which is desirable for chiral molecules sensing. Further, it was demonstrated that the volume-averaged optical chirality could be further increased by changing the number of stacked layers. The results indicated offer a new approach for enhancing chiral fields with metamaterials.

In addition to negative-index materials, all-dielectric metamaterials have recently attracted much attention due to their low heat generation, low intrinsic loss, high-quality factors, and high magnetic resonance[17, 42, 43] Based on these properties, Numerous studies were conducted on these dielectric materials to obtain unique electromagnetic properties. Ho et al. demonstrated that magnetic multipolar resonances of sub-micron silicon spheres are excited by CPL,[19] and the maximum optical chirality and Kuhn's dissymmetry factor are observed near the magnetic resonances. Solomon et al. theoretically investigated the design rules of achieving large area, uniform sign chiral fields with high index dielectric metasurfaces.[44] With matesurfaces composed of silicon disks they obtained a 138-fold enhancement of optical chirality and a 30-fold enhancement of volume-averaged optical chirality near the disks. By tuning the aspect ratio of the disks, the maximum one-handed local chiral field is achieved when the electric and magnetic modes at the same frequency (Fig. 3b) due to an overlap of the electric and magnetic fields in space.

Mohammadi et al. considered similar design rules and obtained highly enhanced chiral near fields with dielectric nanoparticles with the synergistic combination of electric and magnetic resonances.[17] A dielectric dimer composed of two silicon spheres was used (Fig. 3c), and parallel, spatially overlapping, and ideal phase retardation ($\pi/2$) of the electric and magnetic fields are generated in the hot spots. To simplify, holey silicon disk resonators were proposed based on Kerker effect. Uniform chiral near fields were achieved by the periodic resonators. Furthermore, Du et al. used a single hollow silicon disk (Fig. 3d) and systematically analyzed the relationship between the dipolar interference and chiral near-fields using the multipole decomposition method.[21] The results showed that the maximum enhancement of the averaged optical chirality is related to the magnetic dipole and anapole resonances. Meanwhile, the optical chirality enhancement could be tuned by size scaling. This research provided a potential method for chirality detection with all-dielectric metamaterials.

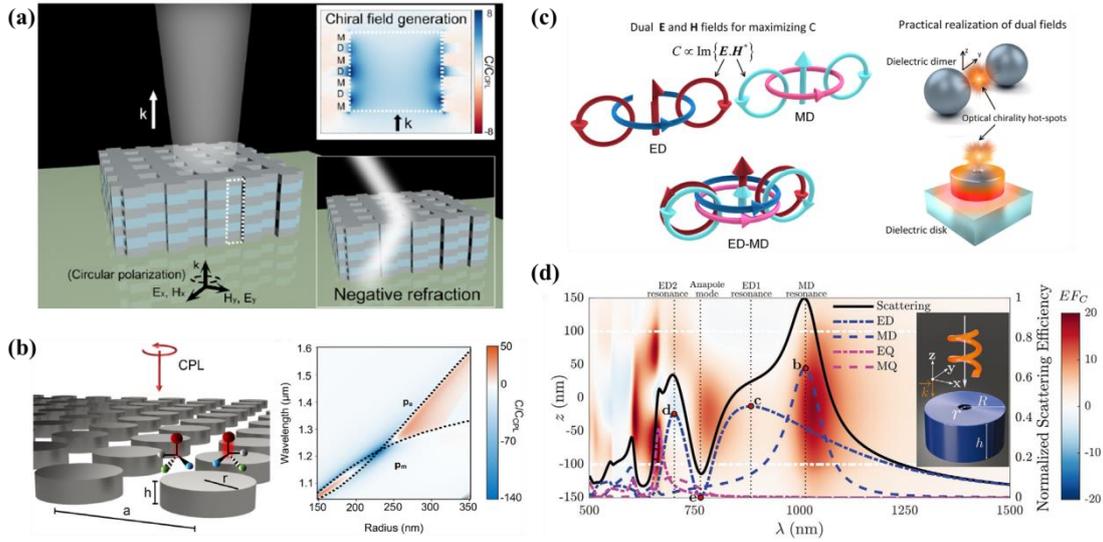

**Figure 3.** (a) Schematic of achiral negative-index metamaterials with globally enhanced chiral fields.[41] This figure has been adapted from ref. 41 with permission from American Physical Society, copyright 2014. (b) Schematic of silicon disk array and the rule of chiral fields with electric and magnetic fields.[44] This figure has been adapted from ref. 44 with permission from American Chemical Society, copyright 2019. (c) Schematic of electric and magnetic dipoles and models with overlapping electric and magnetic dipoles.[17] This figure has been adapted from ref. 17 with permission from American Chemical Society, copyright 2019. (d) Chiral fields and scattering efficiency and the corresponding models of electric and magnetic resonances of a hollow silicon disk.[21] This figure has been adapted from ref. 21 with permission from Wiley-VCH, copyright 2021.

## 5. Ultra-limit detection with chiral near-fields

Due to their excellent photoelectric properties, chiral fields have potential applications in molecular detection, recognition, separation, and sensing. Since the chiral response of natural chiral molecules is very weak, the detection sample is limited to the microgram level. To detect the chiral molecules at the pictogram level, superchiral fields generated by plasmonic metamaterials have been introduced. Hendry et al. first experimentally demonstrated that superchiral fields could be generated in planar gammadion-shaped nanostructures,[9] as shown in **Fig. 4**a. When chiral analytes were adsorbed to the chiral nanostructures, there are obvious wavelength shifts for the chiral responses of nanostructures, which depend on the handedness of the chiral molecules (Fig. 4a (i)). The authors utilized the dissymmetry factor ($g = \frac{n_R - n_L}{n_R + n_L}$, $n_R$, and $n_L$ are the effective refractive indices of the chiral molecules for the right-/left-handed nanostructures, respectively) to measure the strength of interaction between the chiral molecules and the nanostructures. It was $10^6$ times greater than that obtained of detecting the same molecules in solution with CPL. The authors further verified that the strong chiral response was attributed to an increase in the local fields and the superchiral fields caused by the coupling between the different branches of the nanostructures, as shown in Fig 4a (ii).

From another perspective, Tang et al. proposed enantioselective excitation of the chiral molecules with superchiral light, which is produced by the superposition of CPL with opposite chirality and slightly different amplitudes propagating in opposite directions.[7] When the chiral fluorescent compounds are excited by the superchiral light in the experiment, the dissymmetry factors are 11 times greater than those obtained by CPL, and their signs are the opposite for the enantiomers. They pointed out that the enhancement of the dissymmetry factors resulted from the suppression of electric energy density at the nodes. The enantioselective excitation could be carried out for any small chiral molecule.

Conventional chiroptical spectroscopic methods are sensitive to changes in the lower-order structure but not the higher-order structure. However, the superchiral evanescent near fields of the chiral nanostructures are sensitive to the higher-order structure. Based on this, Tullius et al.[45] demonstrated that the higher-order structures of biomolecules could be detected by new label-free biophysical measurements using the superchiral evanescent near fields of "shuriken" nanostructures (Fig. 4b).

It is well known that chiral plasmonic nanostructures could produce large CD signals, which are stronger than those of chiral molecules, preventing the detection of the molecular signals. To solve the problem, Garcia-Guirado et al. developed a plasmonic sensor consisting of a racemic plasmonic arrays of 2D chiral structures, which show strong chiral near-field responses but no far-field CD responses (Fig. 4c).[33] But the CD signal separates from zero by combining a chiral molecular layer. This sensor was used to distinguish the handedness of phenylalanine molecules. Similar racemic nanograting was also used for detecting chiral molecules.

As discussed above, chiral near-fields have a wide range of applications in detection, sensing, and related fields. Therefore, it is important to obtain accurate measurements of chiral near fields around nanostructures. Meinzer et al. experimentally analyzed a chiral near field using luminescence enhancement of achiral dye molecules around chiral nanostructures.[30] As shown in Fig. 4d, the dissymmetry of the luminescence enhancement is correlated with the optical chirality C. This method can accurately detect a chiral near field around the nanostructure, even with a very weak CD response.

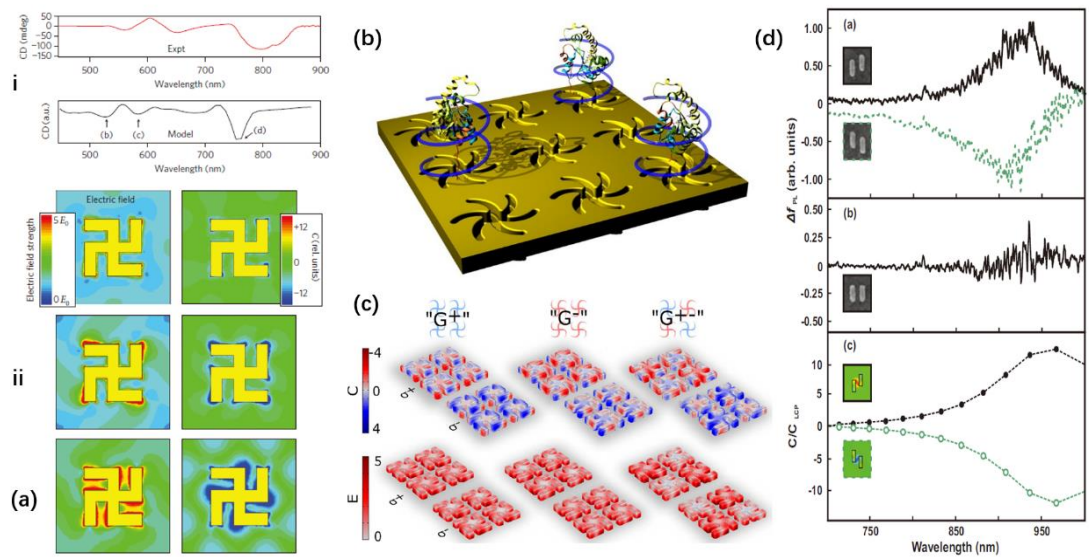

**Figure 4.** (a-i) The CD spectra obtained from an experiment and simulation of planar chiral metamaterials (PCMs). (a-ii) the electric and chiral near-fields distributions near the PCMs.[9] This figure has been adapted from ref. 9 with permission from Nature Publishing Group, copyright 2010. (b) Schematic of the interaction between the chiral nanostructure and molecules.[45] This figure has been adapted from ref. 45 with permission from American Chemical Society, copyright 2015. (c) The electric and chiral near-fields distributions of the chiral and racemic plasmonic arrays.[33] This figure has been adapted from ref. 33 with permission from American Chemical Society, copyright 2018. (d) Dissymmetry between the left-handed and right-handed PL enhancement for the chiral and achiral arrays, and the optical chirality enhancement at the center of the nanorod dimer.[30] This figure has been adapted from ref. 30 with permission from American Physical Society, copyright 2013.

Moreover, strong and uniform chiral near-fields are required to detect a small quantity of molecules or a single molecule. Meanwhile, nanostructures producing chiral near-fields should have simple configurations to be fabricated conveniently and economically. In addition, it is preferable to use linearly polarized light rather than CPL, which requires a broadband circular polarizer. Theoretical research has been conducted to meet these requirements. In practice, molecules are typically randomly distributed around a nanostructureor in a specific area. In 2015, Finazzi et al. pointed out that the volume-averaged optical chirality around small plasmonic nanoparticles could not be larger than that in a plane wave due to the quasistatic nature.[46] The volume-averaged optical chirality may be stronger near larger nanostructures[10] if the signs of the local optical chirality do not change. It was found that one-handed chiral near-fields could be obtained by adding a layer to cover the two opposite corners of squares to prevent cancellation between the chiral near-fields with different handedness in a nanoantenna application (**Fig. 5**a).[12] Thus, the molecules could be placed in the groove and interact with the one-handed chiral near fields. As mentioned at section 3, strong uniform chiral near-fields could be produced by a helical spring structure. Based on this, Schäferling et al. designed a simplified nanostructure composed of a diagonal slit array in a metallic film on a mirror to minimize fabrication difficulty (Fig. 5b).[47] The properties of the chiral fields could be precisely controlled by the distance between the two layers, and one-handed chiral fields could be obtained if the distance is sufficiently small. In experiment, chiral molecules can be placed in the slits and analyzed using chiroptical spectroscopy. Meanwhile, the detection of ROA of a single molecule or a few molecules has attracted significant attention. This process is difficult due to the small cross-section. Hu et al. theoretically proposed an equation to calculate the ROA enhancement factor similar to that of surface-enhanced Raman scattering (SERS). They discussed the ultimate limit of SE-ROA.[36] With the structure consisting of Au block dimer on an Au film, both SERS and ROA are extremely enhanced in the gap, which provides a theoretical basis for realizing the ROA detection of a single molecule in experiments (Fig. 5c).

However, SE-ROA has several drawbacks, such as high photothermal heat and an inability to transfer and enhance the optical chirality from the far field to the near field. Therefore, Xiao et al. theoretically and experimentally demonstrated that the chiral-field enhanced ROA could be significantly increased by tuning the chiral field with a silicon nanodisk array in the dark mode.[48] The average enhancement factor could rearch ~$10^2$ in the near-field region (Fig. 5d).

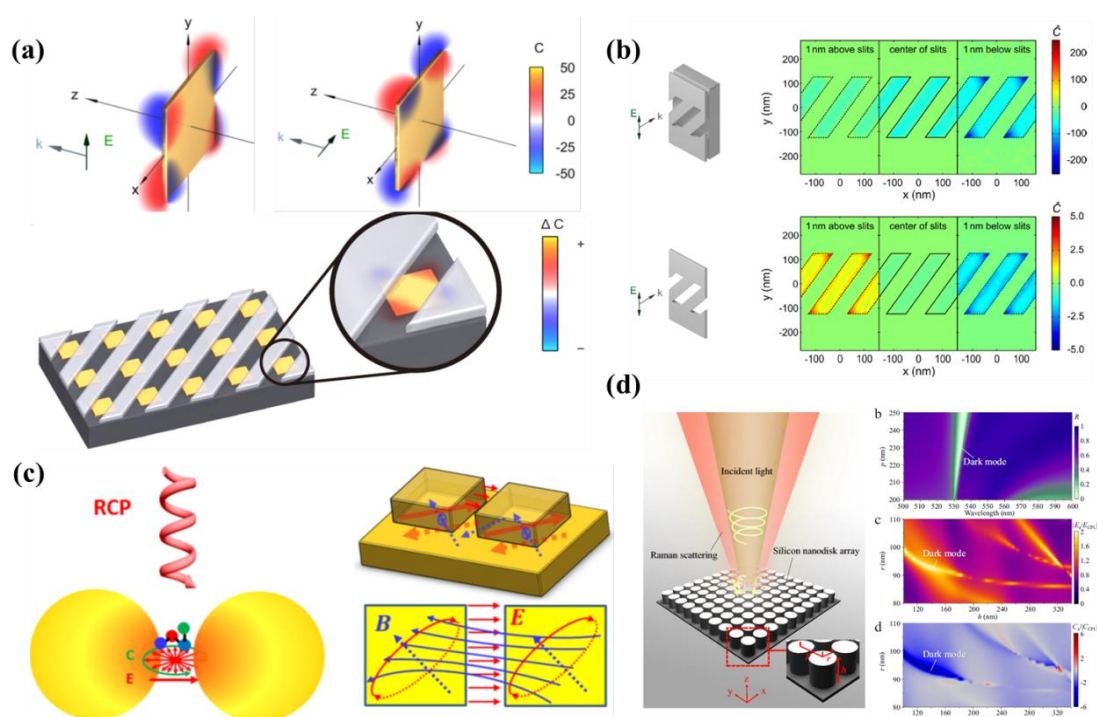

**Figure 5.** (a) The optical chirality distributions of a square under linearly polarized light and the schematic of nanostructures of one-handed chiral near-fields.[12] This figure has been adapted from ref. 12 with permission from The Optical Society, copyright 2012. (b) Chiral near-fields distributions of a diagonal-slit structure with and without a mirror layer.[47] This figure has been adapted from ref. 47 with permission from American Chemical Society, copyright 2016. (c) Schematic of chiral near field generation of a block dimer-film nanostructure under linearly polarized light.[36] This figure has been adapted from ref. 36 with permission from American Chemical Society, copyright 2018. (d) Schematic of a silicon nanodisk array and the far-field and near-field responses.[48] This figure has been adapted from ref. 48 with permission from Nature Publishing Group, copyright 2021, under the Creative Commons Attribution 4.0 License.

## 6. Spin-orbit interaction in chiral fields

As discussed in Section 2, photons have angular momentum, i.e., spin angular momentum and orbital angular momentum. Opposite spin angular momenta correspond to RCP and LCP, whereas the orbital angular momentum is associated with the direction of the wave vector and the phase fronts.[49] The corresponding canonical momentum densities of the field are $P_o = \frac{g}{2} Im[\varepsilon \boldsymbol{E}^* \cdot (\nabla)\boldsymbol{E} + \mu \boldsymbol{H}^* \cdot (\nabla)\boldsymbol{H}]$ and $S_o = \frac{g}{2} Im[\varepsilon \boldsymbol{E}^* \cdot \times \boldsymbol{E} + \mu \boldsymbol{H}^* \cdot \times \boldsymbol{H}]$ where $g = (8\pi\omega)^{-1}$. When light interacts with materials with an inhomogeneous refraction index, the conservation of the optical angular momentum leads to a momentum transfer between the spin (polarization) and orbital (propagation and phase) momenta. As a result, a change in the light's trajectory accompanying a transverse spin-split of the beam occurs.[50, 51] This phenomenon is the photonic spin Hall effect (PSHE), a counterpart of the well-known electronic SHE, where the electrons spin-dependently accumulate at opposite sides of an electron flow due to the spin-orbit interaction.[49]

In conventional cases, the photonic spin-orbit interaction (PSOI) is typically very weak for a small interaction volume. Special techniques are required to observe this phenomenon, such as weak quantum measurements or multiple total internal reflections,[52] limiting its application. Metal nanostructures provide a reliable approach to achieving strong light-matter interaction due to their plasmonic effects. Moreover, a large gradient of permittivity exists on the metal-dielectric interface, resulting in strong PSOI.[53] Yin et al. utilized a metasurface with V-shaped gold antennas to obtain strong PSOI. They directly observed a strong PSHE with a large opposite transverse motion of RCP and LCP represented by the red and blue areas (color which are also represent the chiral field C) in **Fig. 6**a.[51] Subsequently, researchers developed different types of novel spin Hall devices, such as a spin-dependent splitter, Stokes parameter detector, and the Pancharatnam-Berry phase lens, by manipulating the PSOI with subtly designed metasurfaces.[49, 54-56]

Light with both spin and orbital angular momenta can be tailored by the PSOI from the incident light with only spin angular momentum. The interaction between this type of light (Laguerre-Gaussian (LG) beam) and plasmonic nanostructures has been investigated in recent years. The PSOI induces a transformation of the spin and orbital angular momenta. Due to the spatial distribution of the phase front, the isophase surface of the LG beam (with nonzero orbital angular momentum) will sweep over the nanostructure over time. The partial spatial excitation of the plasmonic nanostructure generates strong dichroism that is two orders of magnitude higher than that excited by light with only spin angular momentum (CPL). The plasmon modes excited by light with orbital angular momentum also have significantly different surface charge distributions and superchiral near fields (Fig. 6b). When the incident light has spin and orbital angular momenta with the same sign, the superchiral near fields of the excited plasmon modes are more enhanced than those excited by light without orbital angular momentum (considering the very low intensity of the central area of the LG beam). Meanwhile, when the spin and orbital angular momenta have opposite signs, they cancel the superchiral near field.[6] The PSOI can result in a significant enhancement of the superchiral near field under the right conditions of the spin and orbital momenta.

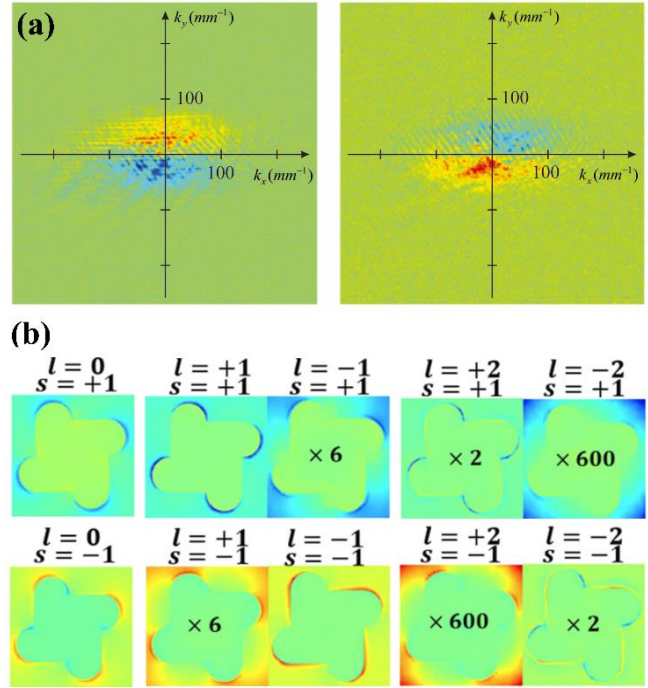

**Figure 6.** (a) A strong PSHE with large opposite transverse motion of refracted right- and left-circularly polarized light near a metasurface with V-shaped gold antennas.[51] This figure has been adapted from ref. 51 with permission from American Association for the Advancement of Science, copyright 2013. (b) Optical chirality of a gold cuboid protuberance chiral structure under a background electric field with different spin and orbital angular momenta. [6] This figure has been adapted from ref. 6 with permission from American Physical Society, copyright 2020.

**7. Chiral field in optical trapping**

Light has linear momentum, which can be transferred to particles via light-matter interaction, pushing the particle forward in the direction of the wave vector. This phenomenon is known as the radiation force, which can be used to manipulate particles.[57] When a particle is located in the narrow beam of a laser, the gradient force occurs due to spatially inhomogeneous intensity. This force has been widely used to power optical tweezers,[58] and is seen as a promising method to interact with chiral molecules due to the effective and noninvasive interaction between chiral light and chiral objects,.[59] In the last decade, the lateral force was discovered to separate chiral objects. It moves objects with different chiralities in opposite directions. The lateral force arises from the transverse spin angular momentum of the field and can be generated by placing an object above a slab, in an evanescent wave, or in the interference light field.[57, 59, 60]

Recently, Li et al. demonstrated the sorting of chiral particles by a tightly focused beam.[61] If a dipolar particle is much smaller than the wavelength, the optical force within the dipole approximation can be analytically expressed as:

$$\langle F \rangle = \nabla \langle U \rangle + \frac{\sigma n_1}{c} \langle S \rangle - \text{Im}[\alpha_{em}] \nabla \times \langle S \rangle - \frac{c\sigma_e}{n_1} \nabla \times \langle L_e \rangle - \frac{c\sigma_m}{n_1} \nabla \times \langle L_m \rangle + \omega \gamma_e \langle L_e \rangle + \omega \gamma_m \langle L_m \rangle + \frac{ck_1^4}{12\pi n_1} \text{Im}[\alpha_{ee} \alpha_{mm}^*] \text{Im}[E \times H^*] \quad (20)$$

with the optical potential[61]

$$\langle U \rangle = \frac{1}{4} \text{Re}[\alpha_{ee}] |E|^2 + \frac{1}{4} \text{Re}[\alpha_{mm}] |H|^2 + \frac{1}{2} \text{Re}[\alpha_{em}] \text{Im}[E \cdot H^*] \quad (21)$$

where $E$ and $H$ are the electric and magnetic fields acting on the particle, $\langle S \rangle$ is the time-averaged Poynting vector, $\langle L_e \rangle$ and $\langle L_m \rangle$ are respectively the electric and magnetic parts of the time-averaged spin angular momentum density, $\text{Im}[E \cdot H^*]$ is the chiral near field, and $n_1$ is the refractive index of the medium with relative permittivity $\varepsilon_1$ and permeability $\mu_1$. $k_1$ is the wavenumber with angular frequency $\omega$; $c$, $\varepsilon_0$, and $\mu_0$ are respectively the light speed, permittivity and permeability in a vacuum; $\alpha_{ee}$, $\alpha_{mm}$, and $\alpha_{em}$ are the electric, magnetic, and chiral polarizabilities of the chiral particle, respectively. $\sigma_e = \frac{k_1 \text{Im}[\alpha_{ee}]}{\varepsilon_1 \varepsilon_0}$, $\sigma_m = \frac{k_1 \text{Im}[\alpha_{mm}]}{\mu_1 \mu_0}$, $\sigma = \sigma_e + \sigma_m - \frac{c^2 k_1^4}{6\pi n_1^2} (\text{Re}[\alpha_{ee} \alpha_{mm}^* + \alpha_{em} \alpha_{em}^*])$, $\gamma_e = 2\omega \text{Im}[\alpha_{em}] - \frac{ck_1^4}{3\pi \varepsilon_1 \varepsilon_0 n_1} \text{Re}[\alpha_{ee} \alpha_{em}^*]$, and $\gamma_m = 2\omega \text{Im}[\alpha_{em}] - \frac{ck_1^4}{3\pi\mu_1\mu_0 n_1}\text{Re}[\alpha_{mm}\alpha_{em}^*]$ are the dimensions of the cross-section. The chirality of the particle can be described by a complex parameter $\kappa$, which is related to $\alpha_{em}$. As shown in **Fig. 7**a, the real part of $\kappa$ induces a radial optical force, while the imaginary part gives rise to a lateral optical force. The radial optical force tends to trap particles with $\kappa > -0.38$ and repels the ones with $\kappa < -0.38$.[61] The lateral optical force has opposite directions for particles with $\kappa$ of different signs (Fig. 7b). Therefore, chiral objects can be sorted.

The radial optical force is a gradient force, which is defined in the first term of equation 20 (Fig. 7c). This term, the optical potential $\langle U \rangle$, is composed of three terms in equation 21. A comparison of equations 12 and 21 shows that the third term $\frac{1}{2}\text{Re}[\alpha_{em}]\text{Im}[\boldsymbol{E} \cdot \boldsymbol{H}^*]$ is proportional to the product of the chirality of the particle and the chiral near field of the beam. We used a model similar to Li's to compare the contribution of the three terms of $\langle U \rangle$ to the radial optical force. The results are shown it in Fig. 7d. The third term ($F_3$) dominates the optical force, and the other two terms ($F_1, F_2$) provide a bias force. This result indicates that the chiral near field of the beam significantly affects the gradient radial optical force. The plasmon-enhanced superchiral near-field can be used to generate a strong chirality-dependent optical force and achieve highly efficient sorting of chiral molecules.

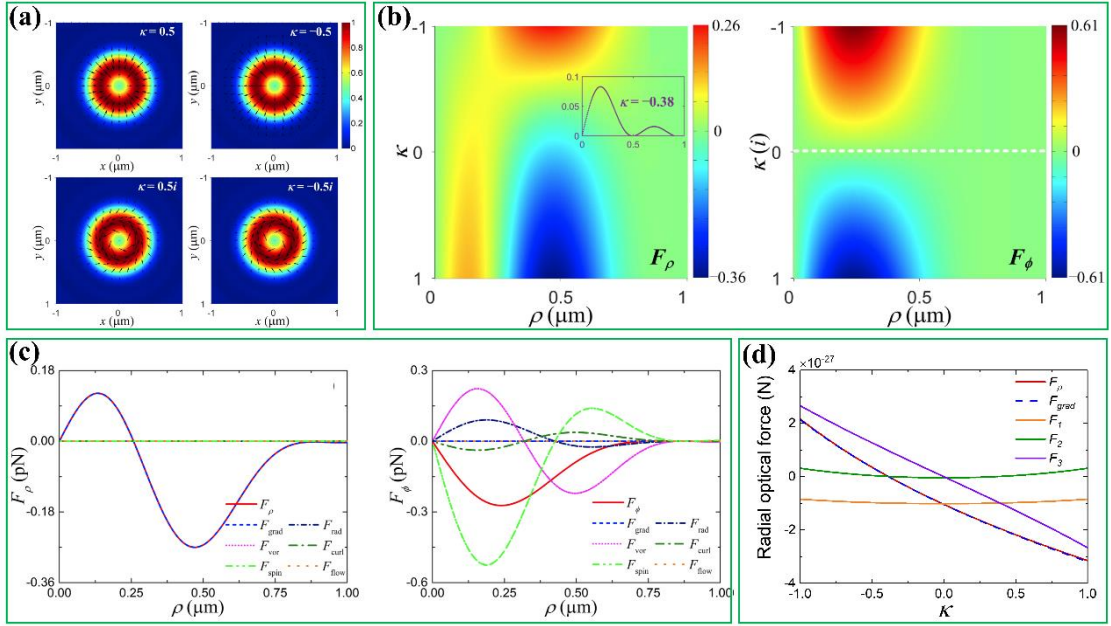

**Figure 7.** (a) Transverse optical force distributions of particles with different chirality parameters.[61] (b) The radial (left) and azimuthal (right) optical forces as functions of the particle's radial displacement and chirality parameter. The inset is a line scan of the radial optical force with $\kappa = -0.38$ in the radial direction.[61] (c) The contributions of the decomposed radial (left) and azimuthal (right) optical forces with $\kappa = 0.5 + 0.5i$.[61] Figures a, b, c have been adapted from ref. 61 with permission from American Physical Society, copyright 2019. (d) The contributions of the decomposed gradient forces with $\rho = 0.1 \ \mu m$.

## 8. Discussion and outlook

Plasmon-enhanced chiral near fields can match chiral molecules and light wavelengths with large size differences, significantly enhancing the CD of the chiral molecules. Plasmonic nanostructures have dimensions of tens to hundreds of nanometers, which is much larger than that of chiral molecules. Thus, it is necessary to develop nanostructures with higher chirality density to achieve higher CD enhancement of the chiral molecules. The fabrication of plasmonic chiral structures with a size of several nanometers or smaller is challenging using current nanofabrication techniques. The plasmonic nanogap enables the generation of a highly distorted chiral near field. However, the small gap size limits the application of this method. The design and fabrication of plasmonic nanostructures to create superchiral near fields with high chirality density are important research topics.

A crucial aspect of research on the plasmon-enhanced chiral near field is its distribution. However, the detection of the near-field distribution is challenging due to a lack of convenient and widely used far-field methods. Luminescence enhancement with dye molecules or quantum dots and leakage radiation microscopy are typically used to image the near field distribution of plasmon waves.[52, 62] However, these methods are not applicable to the detection of plasmon-enhanced chiral near fields due to the low resolution. At present, the most effective method to achieve this goal is scanning near-field optical microscopy (SNOM).[52] It precisely images the amplitude distribution of the chiral near field and provides phase information. Nevertheless, the setup is quite complex and difficult to use, making the detection of the plasmon-enhanced chiral

**Acknowledgements**

This research was supported by the National Natural Science Foundation of China (NSFC) (12074054, 11804035) and the Fundamental Research Funds for the Central Universities (DUT21LK06).

Received: ((will be filled in by the editorial staff))
Revised: ((will be filled in by the editorial staff))
Published online: ((will be filled in by the editorial staff))

Chiral near fields build a bridge to match the chiral molecules and light wavelengths. Employing localized surface plasmons, it is effective to generate and manipulate the highly enhanced superchiral near field, thus strong circular dichroism signals can be obtained. This review presents the principles and state-of-the-art studies on the properties and applications of plasmonic and photonic enhanced chiral near fields.

L. Hu, Z. Sun, Y. Nie, Y. Huang, Y. Fang*

**Plasmonic and photonic enhancement of chiral near fields**

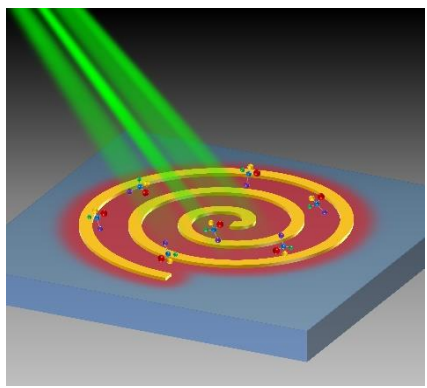